\documentclass[preprint]{aastex}
\newcounter{address}
\newcommand{\latin}[1]{{#1}}
\newcommand{\ie}{\latin{i.e.}}
\newcommand{\eg}{\latin{e.g.}}

\newcommand{\deltacylinder}{\delta_{1\times 8}}

\begin{document}

\title{The dependence on environment of
       the color--magnitude relation of galaxies}
\author{
  David~W.~Hogg\altaffilmark{\ref{NYU},\ref{email}},
  Michael~R.~Blanton\altaffilmark{\ref{NYU}},
  Jarle~Brinchmann\altaffilmark{\ref{MPA}},
  Daniel~J.~Eisenstein\altaffilmark{\ref{Steward}},
  David~J.~Schlegel\altaffilmark{\ref{Princeton}},
  James~E.~Gunn\altaffilmark{\ref{Princeton}},
  Timothy~A.~McKay\altaffilmark{\ref{UMich}},
  Hans-Walter~Rix\altaffilmark{\ref{MPIA}},
  Neta~A.~Bahcall\altaffilmark{\ref{Princeton}},
  J.~Brinkmann\altaffilmark{\ref{APO}},
  \&
  Avery~Meiksin\altaffilmark{\ref{ROE}}
}
\setcounter{address}{1}
\altaffiltext{\theaddress}{\stepcounter{address}\label{NYU}
Center for Cosmology and Particle Physics, Department of Physics, New
York University, 4 Washington Pl, New York, NY 10003}
\altaffiltext{\theaddress}{\stepcounter{address}\label{email}
\texttt{david.hogg@nyu.edu}}
\altaffiltext{\theaddress}{\stepcounter{address}\label{MPA}
Max-Planck-Institut f\"ur Astrophysik,
Karl-Schwarzschild-Str 1, Postfach 1317, D-85741 Garching, Germany}
\altaffiltext{\theaddress}{\stepcounter{address}\label{Steward}
Steward Observatory, 933 N Cherry Ave, Tucson, AZ 85721}
\altaffiltext{\theaddress}{\stepcounter{address}\label{Princeton}
Princeton University Observatory, Princeton, NJ 08544}
\altaffiltext{\theaddress}{\stepcounter{address}\label{UMich}
Department of Physics, University of Michigan,
Randall Lab, 500 East University, Ann Arbor, MI 48109}
\altaffiltext{\theaddress}{\stepcounter{address}\label{MPIA}
Max-Planck-Institut f\"ur Astronomie,
K\"onigstuhl 17, 69117 Heidelberg, Germany}
\altaffiltext{\theaddress}{\stepcounter{address}\label{APO}
Apache Point Observatory, 2001 Apache Point Road,
PO Box 59, Sunspot, NM 88349-0059}
\altaffiltext{\theaddress}{\stepcounter{address}\label{ROE}
Institute for Astronomy, University of Edinburgh, Royal Observatory,
Edinburgh EH9 3HJ, UK}

\begin{abstract}
The distribution in color and absolute magnitude is presented for
55,158 galaxies taken from the Sloan Digital Sky Survey in the
redshift range $0.08<z<0.12$, as a function of galaxy number
overdensity in a line-of-sight cylinder of transverse radius
$1~h^{-1}\,\mathrm{Mpc}$.  In all environments, bulge-dominated
galaxies (defined to be those with radial profiles best fit with large
S\'ersic indices) form a narrow, well defined color--magnitude
relation.  Although the most luminous galaxies reside preferentially
in the highest density regions, there is only a barely detectable
variation in the color (zero-point) or slope of the color--magnitude
relation ($<0.02~\mathrm{mag}$ in $^{0.1}[g-r]$ or $[B-V]$).  These
results constrain variations with environmental density in the ages or
metallicities of typical bulge-dominated galaxies to be under
20~percent.
\end{abstract}


\section{Introduction}

Red, bulge-dominated galaxies show many precise regularities,
including narrow, mass-dependent distributions of colors,
surface-brightnesses, radial profile shapes, ages, chemical
abundances, velocity dispersions, and mass-to-light ratios
\citep[\eg,][]{baum59, faber73a, faber76a, visvanathan77,
djorgovski87, dressler87, kormendy89a, roberts94a, burstein97a,
terlevich01a, eisenstein03b, bernardi03b, bernardi03c, bernardi03d,
blanton03d}.  The most massive of these bulge-dominated galaxies
reside preferentially in the most dense environments, and especially
in rich clusters of galaxies \citep{dressler80a, postman84a}; indeed
the most massive red galaxy populations have the highest average
environmental overdensities \citep{hogg03b}.

Are these trends, \ie, the relationships between mass and environment
and between mass and everything else, related?  In this
\textsl{Letter} we look at the color--magnitude relationship of red,
bulge-dominated galaxies as a function of environment.

Generically, in models in which structure grows gravitationally, the
most dense regions of the Universe will have collapsed earlier and
will contain the most massive objects.  Detailed modeling of galaxy
formation in current cosmological simulations bears this out; galaxies
in very high density regions are predicted to be more luminous and
more red (older) than those in typical density regions
\citep[\eg,][]{cen93b, kauffmann93a, kauffmann96a, kauffmann99a,
benson00a, diaferio01a}.  While these models predict morpholgy, color,
and luminosity variations with overdensity, it is not clear whether
they predict color differences at fixed morphology and luminosity.

The Sloan Digital Sky Survey (SDSS) is the best available data set for
work on these questions, because of its sample size, high
signal-to-noise imaging, sky coverage, and complete spectroscopy
\citep[\eg,][]{york00a}.  Indeed, the SDSS has already made some
progress on related issues, including the dependence of clustering on
luminosity and color \citep{zehavi02a}, the star-formation rate as a
function of environment \citep{gomez03a}, the mean red-galaxy spectrum
as a function of environment \citep{eisenstein03b}, and the mean
galaxy environment as a function of color and luminosity
\citep{hogg03b, blanton03d}.

In what follows, a cosmological world model with
$(\Omega_\mathrm{M},\Omega_\mathrm{\Lambda})=(0.3,0.7)$ is adopted,
and the Hubble constant is parameterized
$H_0=100\,h~\mathrm{km\,s^{-1}\,Mpc^{-1}}$, for the purposes of
calculating distances and volumes \citep[\eg,][]{hogg99cosm}.

\section{Data}

The SDSS is taking $ugriz$ CCD imaging of $10^4~\mathrm{deg^2}$ of the
Northern Galactic sky, and, from that imaging, selecting $10^6$
targets for spectroscopy, most of them galaxies with
$r<17.77~\mathrm{mag}$ \citep[\eg,][]{gunn98a,york00a,stoughton02a}.
All the data processing: astrometry \citep{pier03a}, source
identification, deblending and photometry \citep{lupton01a},
calibration \citep{fukugita96a,smith02a}, spectroscopic target
selection \citep{eisenstein01a,strauss02a,richards02a}, spectroscopic
fiber placement \citep{blanton03a}, and spectroscopic data reduction
are performed with automated SDSS software.  Redshifts are measured on
the reduced spectra by an automated system, which models each galaxy
spectrum as a linear combination of stellar populations (Schlegel, in
preparation).

To each galaxy's extracted radial profile, a seeing-corrected S\'ersic
model is fit \citep[as in][]{blanton03d}.  The S\'ersic model is that
surface brightness $I$ is proportional to $\exp (-r/r_0)^{1/n}$, where
$r$ is the angular radius from the galaxy center, $r_0$ is a
characteristic radius, and $n$ is a free parameter known as the
``S\'ersic index''.  Galaxies with more concentrated radial profiles
have larger $n$; an exponential disk has $n=1$ and a de~Vaucouleurs
profile has $n=4$.

Galaxy luminosities (absolute magnitudes) and colors \citep[measured
by the standard SDSS petrosian technique;][]{petrosian76a} are
computed in fixed bandpasses, using Galactic extinction corrections
\citep{schlegel98a} and $K$ corrections \citep[computed with
\texttt{kcorrect v1\_11};][]{blanton03b}.  Galaxy magnitudes are $K$
corrected not to the redshift $z=0$ observed bandpasses but to bluer
bandpasses $^{0.1}g$, $^{0.1}r$ and $^{0.1}i$ ``made'' by shifting the
SDSS $g$, $r$, and $i$ bandpasses to shorter wavelengths by a factor
of 1.1 \citep[as in][]{blanton03b, blanton03d}.  This means that
galaxies at redshift $z=0.1$ have trivial \citep[but not
zero;][]{hogg02c} $K$ corrections.  The full error analysis of SDSS
photometry is not complete, but scatter in the $g$, $r$, and $i$ band
magnitudes appears to be better than $0.03~\mathrm{mag}$ (RMS).

For a red galaxy, $^{0.1}[g-r]$ (calibrated to the AB magnitude
system) can be converted approximately to $[B-V]$ with
\begin{eqnarray}\displaystyle
\left[B-V\right]_\mathrm{AB} &=& 0.83\,\,^{0.1}\left[g-r\right]
                      - \left[0.025~\mathrm{mag}\right] \nonumber\\
\left[B-V\right]_\mathrm{AB} &=& \left[B-V\right]_\mathrm{Vega}
                      - \left[0.115~\mathrm{mag}\right] \quad ,
\end{eqnarray}
where the subscripts indicate AB or Vega-relative calibration.  This
transformation is robust because the $^{0.1}g$ and $^{0.1}r$
bandpasses are very close to $B$ and $V$ in effective wavelength.

The sample of galaxies used here is a subset of an SDSS statistical
sample known as ``NYU LSS \texttt{sample12}'', further selected to
have apparent magnitude in the range $14.5<r<17.77~\mathrm{mag}$, and
redshift in the narrow range $0.08<z<0.12$.  These cuts left 55,158
galaxies.

The sample is statistically complete, with small incompletenesses
coming primarily from (1) galaxies missed because of mechanical
spectrograph constraints \citep[$\sim 7$~percent;][]{blanton03a},
which does lead to a slight under-representation of high-density
regions, and (2) spectra in which the redshift is either incorrect or
impossible to determine ($<1$~percent).  In addition, there are some
galaxies ($\sim 1$~percent) blotted out by bright Galactic stars, but
this incompleteness should be uncorrelated with galaxy properties.

For each galaxy in the sample there is a computed inverse selection
volume $1/V_\mathrm{max}$, where $V_\mathrm{max}$ is the total
comoving volume in which the galaxy could lie and still be included
in the sample, accounting for the flux limits, surface brightness
limits, redshift limits, and completeness as a function of angle
\citep[as in][]{blanton03d}.

We employ an environment overdensity estimate $\deltacylinder$ based
on the SDSS spectroscopic sample; it is a measure of the
three-dimensional redshift-angle space number density excess around
each galaxy.  The comoving transverse distances and comoving
line-of-sight distances \citep[\eg,][]{hogg99cosm} are computed
between each spectroscopic galaxy and its neighboring spectroscopic
galaxies (not attempting to correct for peculiar velocities).  For
each galaxy, neighbors within a cylinder with a transverse comoving
radius of $1\,h^{-1}~\mathrm{Mpc}$ and a comoving half-length of
$8\,h^{-1}~\mathrm{Mpc}$ in this space are counted; the result is
divided by the no-clustering prediction made from the galaxy
luminosity function \citep{blanton03c} and the sample selection
critera, and unity is subtracted to produce the overdensity estimate
$\deltacylinder$.  A galaxy in an environment with the cosmic mean
density has $\deltacylinder=0$.  The statistical properties of the
estimator $\deltacylinder$ vary with redshift because the density of
the sample varies with redshift, but the sample considered here is
selected to be only in the narrow redshift interval $0.08<z<0.12$.  It
is worthy of note that even galaxies with $\deltacylinder >50$ (the
highest density subsample used below) are not found exclusively inside
the cores of rich clusters.

\section{Results}

\begin{figure}
\plotone{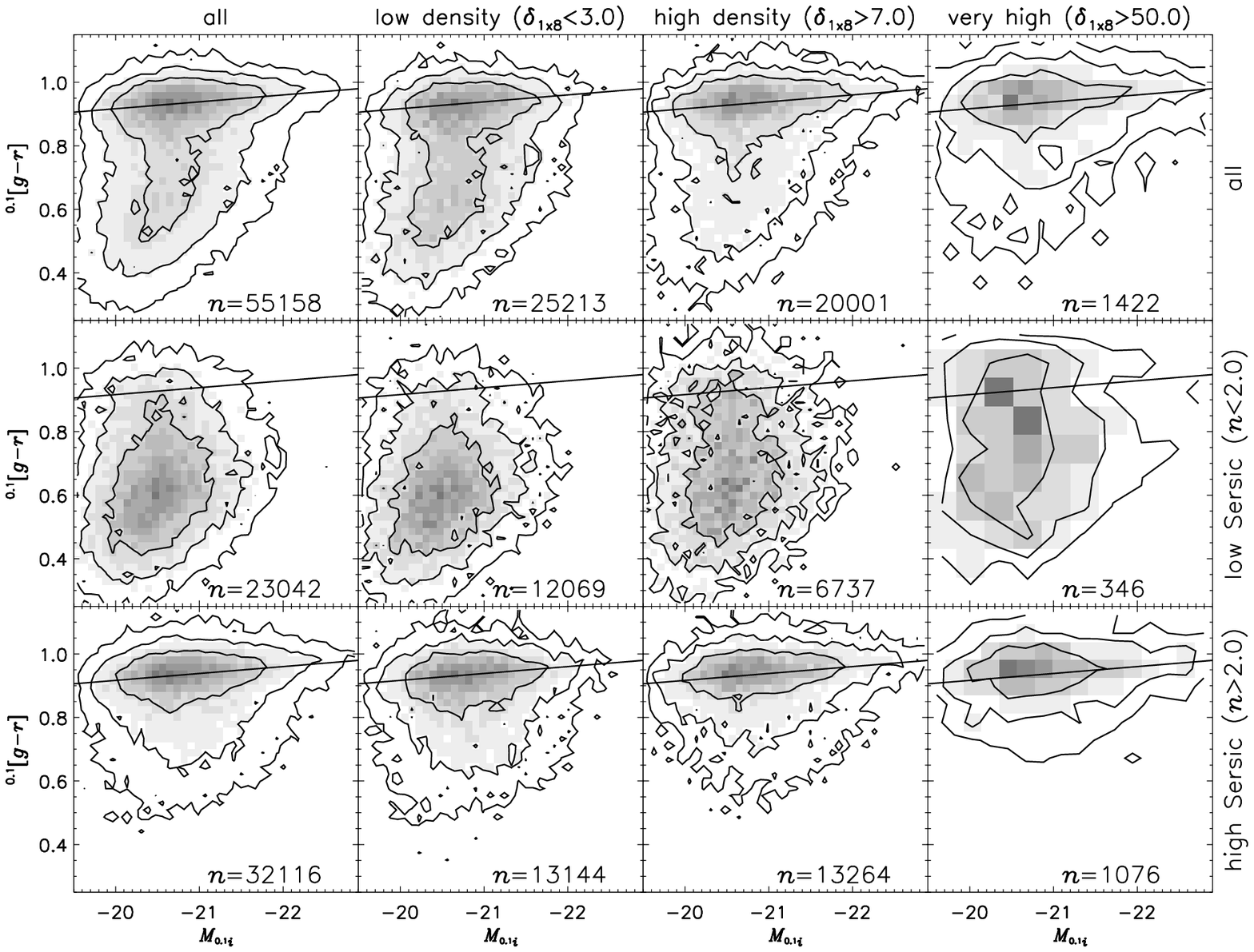}
\caption[]{The top left panel shows the distribution in absolute
magnitude $M_{^{0.1}i}$ in the $^{0.1}i$ band and $^{0.1}[g-r]$ color
for the entire sample.  The columns show subsamples cut in overdensity
$\deltacylinder$ (a mean density environment has $\deltacylinder=0$).
The rows show subsamples cut in S\'ersic index $n$ (an exponential
disk has $n=1$ and a de~Vaucouleurs profile has $n=4$).  In each
panel, the greyscale monotonically represents the abundance of sample
galaxies in the two-dimensional space of color and magnitude and the
contours enclose 52.0, 84.3, and 96.6~percent of the sample.  This
represents properties of the sample; the data have not been weighted
by $1/V_\mathrm{max}$.  Overplotted on all panels is the same
straight, solid line showing the best-fit color--magnitude relation
from the lower-left panel, fit as described in the
text.\label{fig:red_sequence}}
\end{figure}

Figure~\ref{fig:red_sequence} shows the distribution in color and
magnitude for the entire sample, and subsamples chosen to lie in low
density regions ($\deltacylinder <3.0$), high density regions
($\deltacylinder >7.0$), and very high density regions
($\deltacylinder >50$), and subsamples chosen to have low S\'ersic
index ($n<2.0$) and high S\'ersic index ($n>2.0$).  Overplotted on all
panels is the same color--magnitude relation, a linear fit to the
dependence of color on magnitude for the all-environment, high
S\'ersic subsample, with sigma-clipping at $2\,\sigma$ in the color
direction.  The apparent bimodality of the galaxy population in
Figure~\ref{fig:red_sequence} and the value of the S\'ersic index for
separating populations has been made evident previously
\citep{blanton03d}.

The distribution of galaxies is different in the different subsamples.
The most striking difference is that the bulge-dominated ($n>2.0$)
galaxies are much redder than the disk-dominated galaxies.  Another
striking difference is that the bulge-dominated galaxies are
over-represented in the higher density regions.  Another is that the
very most luminous ($M_{^{0.1}i}\sim -22.8~\mathrm{mag}$) galaxies are
over-represented in the higher density regions.

Among the bulge-dominated ($n>2.0$) galaxies it is just as striking
that the color--magnitude relation is independent of the environmental
overdensity.  There are environmental differences in the distribution
of galaxies in the color--magnitude plane, but the most likely color
for a bulge-dominated galaxy of any given luminosity is not much
redder in the highest density environments than it is in the lowest.

\begin{figure}
\plotone{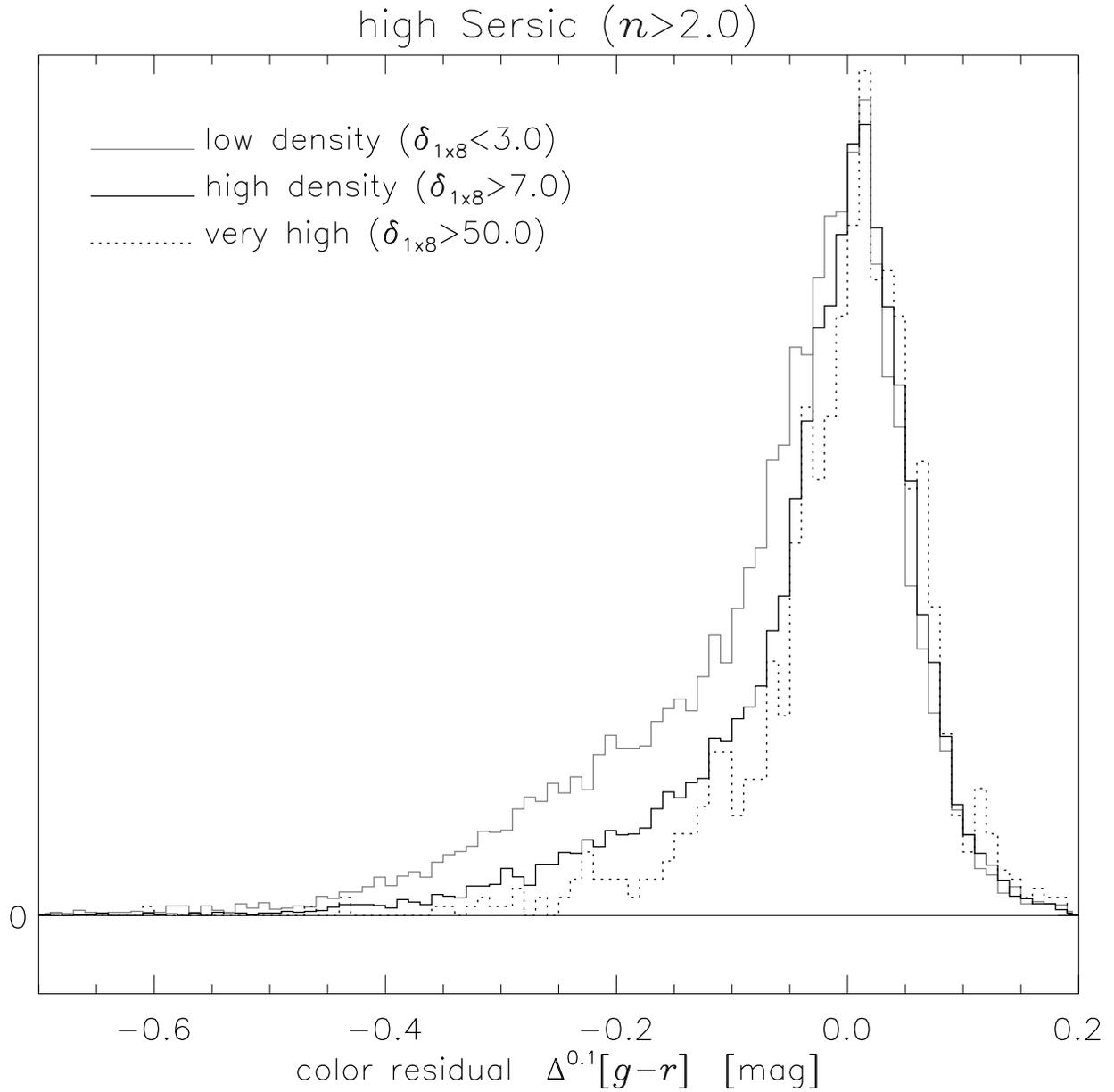}
\caption[]{The distribution of color residuals (observed minus fit)
relative to the linear fit plotted in Figure~\ref{fig:red_sequence}
for the high sersic samples at low and high density.  The histograms
have been arbitrarily re-normalized to have similar peak
heights.\label{fig:red_sequence_hist}}
\end{figure}

Figure~\ref{fig:red_sequence_hist} shows the color-direction residuals
away from the linear fit, for the high S\'ersic-index galaxies in low,
high, and very high density regions.  The distributions differ in the
abundance of blue galaxies (and therefore in the shape of the residual
distribution), as noted previously \citep{pimbblet02a,kuntschner02a}.
All three overdensity subsamples show strong red peaks, with similar
modes and dispersions.  In detail, the highest density subsample shows
the narrowest and reddest peak, with a mode that is slightly redder
(by $<0.02~\mathrm{mag}$ in $^{0.1}[g-r]$) than that of the lowest
density subsample.

This independence of red galaxy colors on environment is consistent
with a previous SDSS result \citep{bernardi03d} which was based on a
restricted sample of early-type galaxies, selected on the basis of
radial profile, ellipticity, color, and lack of emission lines.  Small
variations with environmental density have been found in the mean
strengths of narrow spectral features in red galaxies
\citep{eisenstein03b}, but evidently they are not strong enough to
affect the broad-band colors significantly.

There are several other trends visible in
Figure~\ref{fig:red_sequence}.  The bulge-dominated galaxies show a
larger ``tail'' of blue outliers in lower density regions; some of the
galaxies in this tail are post-starburst galaxies \citep{quintero03a};
some of the others are LINERs (work in preparation).  The high-density
regions contain a small population of red, exponential, sub-$L^\ast$
galaxies \citep[$L^\ast$ is at $M_{^{0.1}i}\approx
-20.8~\mathrm{mag}$,][]{blanton03c} not seen at lower densities.  This
population is visible in previous SDSS work \citep{hogg03b}.

The slope (color over absolute magnitude) of the color-magnitude
relation plotted in Figure~\ref{fig:red_sequence} is
$-0.022~\mathrm{mag\,mag^{-1}}$, and the HWHM of the relation visible
in Figure~\ref{fig:red_sequence_hist} is $<0.05~\mathrm{mag}$
(expressed as a limit because calibration errors may contribute
significantly).  If high luminosity red galaxies are made from the
mergers of lower luminosity galaxies without further star formation,
the color-magnitude relation is difficult to explain.  On the other
hand, the major merger of two equal-luminosity galaxies on the
color--magnitude relation does not, by itself, create a noticeable
outlier in this dataset.

\begin{figure}
\plotone{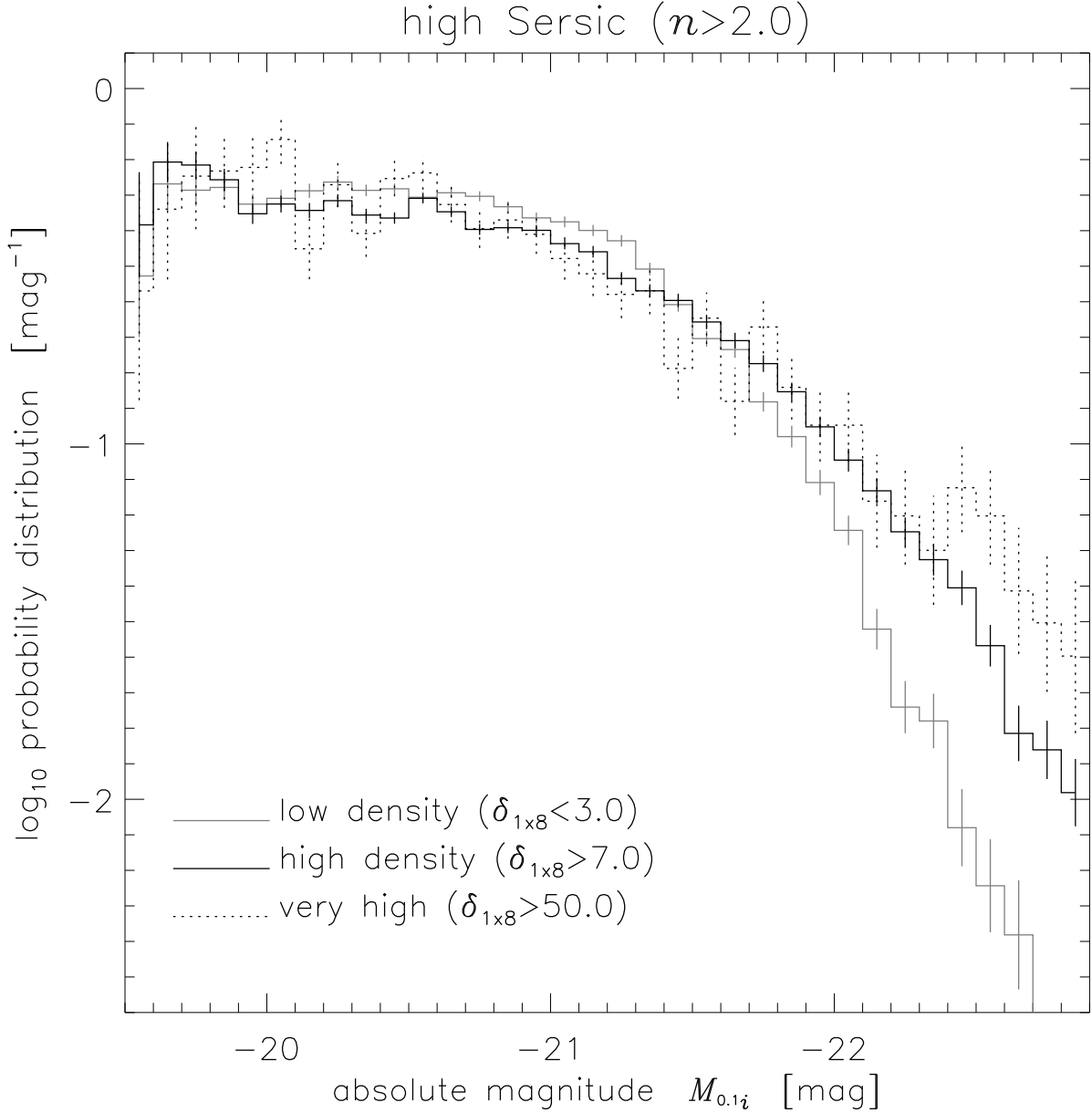}
\caption[]{The $1/V_\mathrm{max}$-weighted distribution function of
galaxy absolute magnitudes for the high S\'ersic ($n>2.0$) galaxies
for the different density environments.  The error bars show Poisson
uncertainties only.\label{fig:red_sequence_lpdf}}
\end{figure}

The fact that the distribution of bulge-dominated galaxy luminosities
is a strong function of environment is emphasized in
Figure~\ref{fig:red_sequence_lpdf}, which shows the luminosity
distributions of the high S\'ersic ($n>2.0$) galaxies in the three
environment subsamples.  The abundance of high luminosity galaxies
increases rapidly with environmental overdensity.  This increase in
average luminosity with increasing overdensity is related to (but not
identical with) the previously shown increase in average overdensity
with increasing luminosity (\ie, the converse relation) for these
galaxies \citep{hogg03b}.

A change of $\sim 0.01~\mathrm{mag}$ in $^{0.1}[g-r]$ corresponds to
$\sim 0.8~\mathrm{Gyr}$ of aging or $\sim 0.04~\mathrm{dex}$ in
metallicity for an old ($\sim 10~\mathrm{Gyr}$), solar metallicity
stellar population \citep[\eg,][]{charlot03a}.  Of course it is
possible that there are compensating effects between metallicity and
age \citep[\eg,][]{kuntschner02a}, although no evidence for such a
trade-off appears in the analysis of SDSS spectra; what appears is
evidence that small age effects, small metallicity effects, and small
relative abundance effects (\eg, alpha enhancement) all come in at
comparable levels \citep{eisenstein03b}.

It is difficult to imagine how high and low density regions
of the Universe can evolve in their different ways without affecting
the color--magnitude relation of the bulge-dominated galaxy
population, in color or slope, by more than what is observed here.

\acknowledgments We thank Ivan Baldry, Mariangela Bernardi, Douglas
Finkbeiner, Amy Kimball, Yeong Loh, Robert Lupton, Bob Nichol, Jim
Peebles, David Wake, and Simon White for useful discussions and
software.  This research made use of the NASA Astrophysics Data
System.  MRB and DWH are partially supported by NASA (grant
NAG5-11669) and NSF (grant PHY-0101738). DJE is supported by NSF
(grant AST-0098577) and by an Alfred P. Sloan Research Fellowship.

Funding for the creation and distribution of the SDSS has been
provided by the Alfred P. Sloan Foundation, the Participating
Institutions, the National Aeronautics and Space Administration, the
National Science Foundation, the U.S. Department of Energy, the
Japanese Monbukagakusho, and the Max Planck Society. The SDSS Web site
is http://www.sdss.org/.

The SDSS is managed by the Astrophysical Research Consortium (ARC) for
the Participating Institutions. The Participating Institutions are The
University of Chicago, Fermilab, the Institute for Advanced Study, the
Japan Participation Group, The Johns Hopkins University, Los Alamos
National Laboratory, the Max-Planck-Institute for Astronomy (MPIA),
the Max-Planck-Institute for Astrophysics (MPA), New Mexico State
University, University of Pittsburgh, Princeton University, the United
States Naval Observatory, and the University of Washington.

\bibliographystyle{apj}
\bibliography{apj-jour,ccpp}

\end{document}